 \documentclass[pmlr,twocolumn]{jmlr} 

\usepackage{booktabs}
\usepackage[load-configurations=version-1]{siunitx} 
\usepackage{mdwtab}

\theorembodyfont{\upshape}
\theoremheaderfont{\scshape}
\theorempostheader{:}
\theoremsep{\newline}

\usepackage[font=small,skip=0pt]{caption}

\jmlrvolume{ML4H Extended Abstract}
\jmlryear{2020}
\jmlrsubmitted{2020}
\jmlrpublished{}
\jmlrworkshop{Machine Learning for Health (ML4H) 2020}

\title[Gender Disparities in Time to Diagnosis]{Exploring Gender Disparities in Time to Diagnosis}

 \author{
 \Name{Tony Y. Sun} \Email{tys2108@cumc.columbia.edu}\\
\Name{Oliver J. {Bear Don't Walk IV}}
\Email{ob2285@cumc.columbia.edu} \\
\Name{Jennifer L. Chen} \Email{jc4686@columbia.edu}\\
\Name{Harry {Reyes Nieva}} \Email{harry.reyes@columbia.edu}\\
\Name{Noémie Elhadad}
\Email{noemie.elhadad@columbia.edu}\\
\addr Columbia University}

\begin{document}
\maketitle
\begin{abstract}
Sex and gender-based healthcare disparities contribute to differences in health outcomes. We focus on time to diagnosis (TTD) by conducting two large-scale, complementary analyses among men and women across 29 phenotypes and 195K patients. We first find that women are consistently more likely to experience a longer TTD than men, even when presenting with the same conditions. We further explore how TTD disparities affect diagnostic performance between genders, both across and persistent to time, by evaluating gender-agnostic disease classifiers across increasing diagnostic information. In both fairness analyses, the diagnostic process favors men over women, contradicting the previous observation that women may demonstrate relevant symptoms earlier than men. These analyses suggest that TTD is an important yet complex aspect when studying gender disparities, and warrants further investigation.
\end{abstract}


\vspace{-16pt}
\section{Introduction}
\label{sec:intro}
\vspace{-7pt}
Sex and gender\footnote{We recognize gender is a social construct while sex is a biological variable~ \citep{WHO2002}. We use the terms ``gender,'' ``men,'' and ``women'' based on available labels.} play significant roles in the initial diagnosis and treatment of disease. An increasing body of literature has documented numerous examples of  biological sex differences in disease presentation as well as gender disparities in care~\citep{seeland2013, regitz-zagrosek2012,westergaard2019}. For example, recent research shows that women with acute myocardial infarction present with different symptoms than men (i.e., sex difference). Healthcare providers, which tend to use diagnostic criteria based on typical male presentation, fail to recognize in time heart attacks in women (i.e., gender bias). These two factors contribute to decreased survival outcomes for women~\citep{shaw2006,mehta2016,greenwood2018patient}.

\vspace{-2pt}
In the machine learning for healthcare community, there is growing interest in creating methodology to identify health disparities as well as to mitigate them~\citep{20chen}. Here, we focus on time to diagnosis (TTD), traditionally defined as the interval from first alert symptoms to diagnosis of a disease in a patient~\citep{launay2016reporting}, and explore potential TTD-related gender disparities. Using the electronic health records of 195,000 patients (113K women and 81K men), and phenotype definitions for 29 conditions (wherein each patient is assigned an official diagnosis time), we carry out two complementary analyses for each of the 29 conditions: \textbf{(1)} we compare mean TTD between men and women; and \textbf{(2)} given a gender-agnostic disease classifier, we assess its gender fairness as presenting symptoms are accumulated over time and holistically assess bias in a time-persistent manner (through mean squared discrimination). This study is approved by our IRB.

\vspace{-17pt}
\section{Dataset}
\label{sec:data}
\vspace{-7pt}
Using data from Columbia University Irving Medical Center, we identified patients at least 13 years of age with at least 3 years of continuous observation between January 2010 and January 2019. Using 29 established phenotype definitions publicly available on the OHDSI platform, we constructed 29 corresponding cohorts ~\citep{OHDSI}. As such, patients in each cohort can be aligned in time according to their official diagnosis time. 

For each patient in a given cohort, we extracted their gender and all coded condition occurrences observed within 3 years prior to phenotype diagnosis. Along with each coded condition, we keep track of date of its first occurrence in a patient longitudinal record. The final dataset comprises 113,885 women (mean age, 55.7 years) and 81,742 men (mean age, 55.9 years) with 13,852 unique condition codes. 
\vspace{-17pt}
\section{Time to Diagnosis among Women and Men}
\label{sec:ttd}
\vspace{-7pt}
\paragraph*{Analysis.}
Here we extend the traditional definition of TTD to include all presenting conditions prior to a phenotype official diagnosis. We compute a given condition's TTD as mean time interval between the condition's first occurrence in a patient record and the phenotype diagnosis time (Fig. \ref{fig:ttd_analysis}).

\begin{figure}
\floatconts
    {fig:ttd_analysis}
    { \setlength{\belowcaptionskip}{-15pt} \caption{TTD computation for each phenotype (e.g., Crohn's disease).}}
    {\includegraphics[width=\columnwidth]{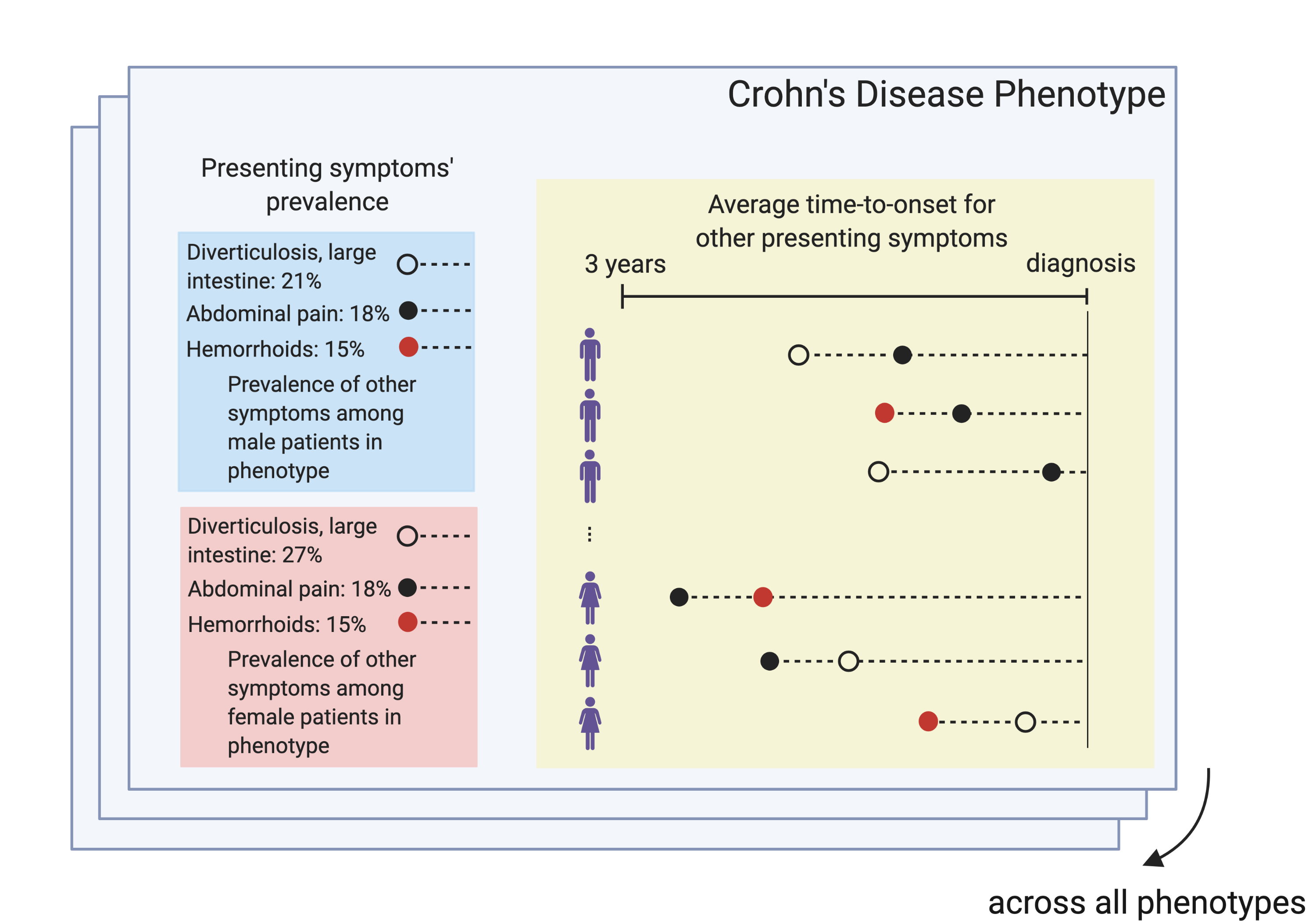}}
\end{figure}

\vspace{-5pt}
\paragraph*{Results.}
 Women were consistently diagnosed later than men, when presenting with the same conditions prior to a phenotype's diagnosis. In $61.5\%$ of all conditions for patients that go on to develop the same phenotype, women were diagnosed later than men (mean difference of $14.8$ days with the same presentation), when averaged across the 29 phenotypes. Additionally, 1,907 conditions ($13.76\%$) had 100+ day TTD difference across gender.

\vspace{-17pt}
\section{Time-Variant Model Fairness in Diagnosis Classification}
\label{sec:tvmfdc} 
\vspace{-7pt}

\paragraph*{Analysis.} For each phenotype, we train a gender agnostic binary classifier that diagnoses patients based on clinical observations documented up to 3 years prior to the official disease diagnosis. As such, the trained classifier is ``fully knowledgeable'' as it contains all observations up to time of diagnosis, particularly the relevant symptoms (e.g., abdominal pain indicating Crohn's disease). (See Appendix~\ref{app:classifier} for details).  

Given that women have on average longer TTD than men, we assess whether women should be diagnosed earlier than men according to the trained classifier. To test this hypothesis, we measure the classifiers' performance on a cohort of patients with decreasing levels of right censoring (see Fig. \ref{fig:bin_figure}) up to 3 years prior to their diagnosis. In essence, this testing procedure aims to mimic a provider diagnosing patients at increasing time steps with knowledge limited to patient history accumulated thus far. 

\begin{figure}
\floatconts
    {fig:bin_figure}
    {\setlength{\belowcaptionskip}{-20pt} \vspace*{-3mm}\caption{Test sets are generated with varying levels of right censoring, mimicking varying amounts of patient history available to providers.}}
    {\includegraphics[width=\columnwidth]{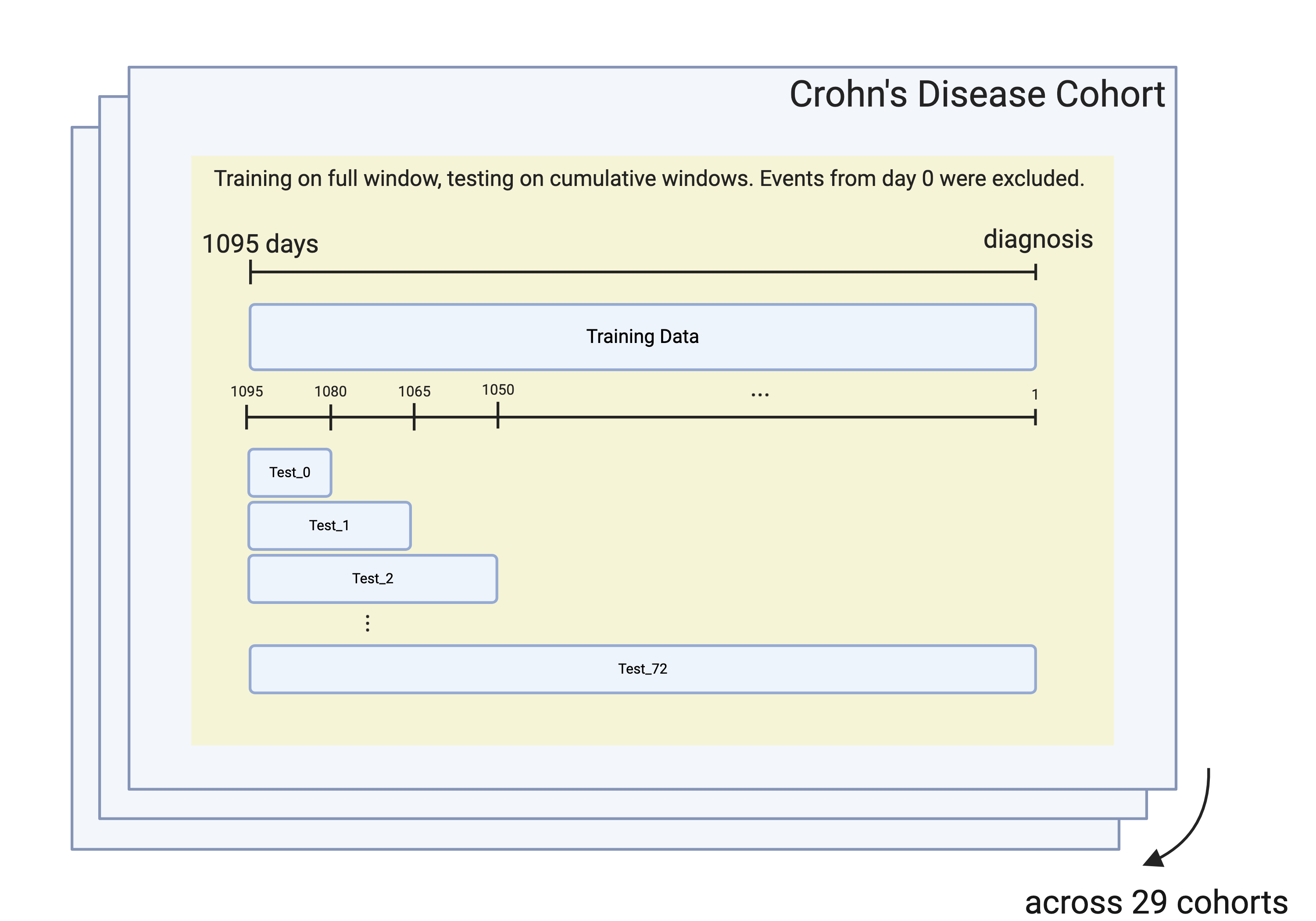}}
\end{figure}

 Formally, our evaluation procedure is as follows. Let $N_d$ be the number of patients for a particular disease $d$, and $C_d$ the number of symptoms (clinical concepts) related to this disease. To evaluate growing windows with right censoring, let $w$ be window size in days, therefore $b = \lceil 1095 / w \rceil$ is the number of possible windows (buckets) of test data. For each window $i \in \{1, \dots, b \}$, we increase available symptoms per patient from day 0 (1095 days before diagnosis) to day $i * w$. Each right-censored test data window $T_i$ is a $N_d \times C_d$ binary sparse matrix where $T^{(p, s)} = 1$ if patient $p$ has symptom $s$ at or before the maximum window date $i * w$. 
 
 Iterating through each test window $T_1, \dots, T_b$, we quantify fairness gaps at each window to understand trends in bias as the diagnostic procedure progresses. In line with established fairness literature, we measure the equality of opportunity of the positive and negative class (recall and specificity gap, respectively) \citep{zhang2020hurtful}, and the precision and accuracy gap at each test window (Table \ref{tab:fairness_gaps}). For example, recall gap is the recall of diagnosing male patients minus the recall of female patients. In diagnosis, recall gap is important as minimizing false negatives ensures patients with a disease are not left untreated. Our approach to analyzing fairness gap over time is extensible to any definition of fairness relevant to the data domain. 
 
 We also sought to quantify gender bias per model persistent across time. \cite{chen2018classifier} proposes that fairness gaps (discrimination level) of a trained model are attributable to the bias, variance, and noise of a dataset conditioned on subgroup attributes. They experimentally suggest that increasing training data size for a dataset can reduce fairness gaps if the discrimination is primarily due to variance within discriminated subgroups. Our analysis of gender bias
 differs by evaluating on the test data in a longitudinal manner, decreasing the amount of right-censoring at each time step and measuring the fairness gap over time. To measure the average fairness gap as information accumulates over time, we leverage mean squared error (MSE) and the overall directionality of bias over $b$ time windows, combining these concepts into mean squared discrimination (MSD). See \ref{apd:fairness} for mathematical explication.

\begin{figure*}[t]
\floatconts
    {fig:mse_recall}
    {\setlength{\belowcaptionskip}{-10pt}\vspace*{-5mm}\caption{Recall MSD across phenotypes. Blue circles indicate better recall for men; pink, for women. Circle size is proportional to the size of the dataset (Table \ref{tab:ap1}). Error bars correspond to 95\% confidence intervals. {\tiny *Excludes non-melanoma skin cancer  }}}
    {\includegraphics[width=0.7\textwidth]{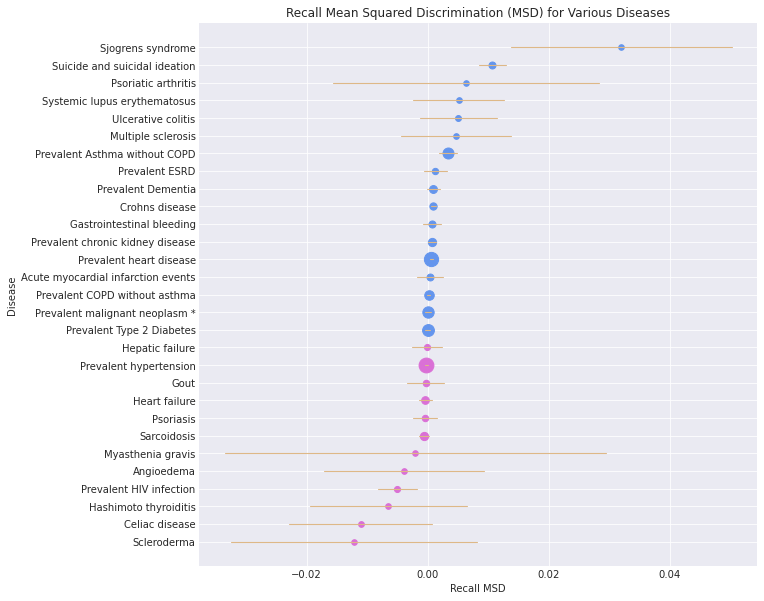}}
\end{figure*}

\vspace{-5pt} 
\paragraph*{Results.}
 Analyzing fairness gaps across time steps for individual phenotypes, such as Crohn's disease, we find that trends in TTD disparity match observed disparity in our previous analysis (section \ref{sec:ttd}). As the amount of patient history increases, the Crohn's disease model's recall for both genders increases, but performance for men is consistently higher than women (Fig. \ref{fig:crohn_gender}). Consistently positive recall gap (indicating better model recall for men) contradicts our previous finding that women present symptoms earlier than men. Fitting the recall gap using linear regression, we observe that male-favoring recall gap decreases $-2.65\mathrm{e}^{-4}$ at each window step (Fig. \ref{fig:crohn_recall_gap}), demonstrating convergence in recall between genders over time.

\begin{figure}[t]
\floatconts
    {fig:crohn_gender}
    { \setlength{\belowcaptionskip}{-25pt} \vspace*{-5mm}\caption{Gender-specific recall at varying levels of right censoring for Crohn's Disease.}}
    {\includegraphics[width=\columnwidth]{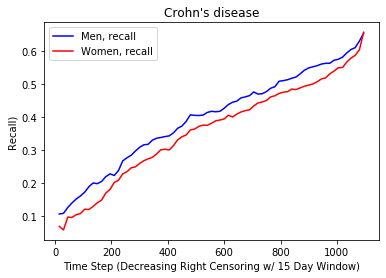}}
\end{figure}

While fine-grained per-phenotype analysis allows us to compare observed fairness trends with hypothesized TTD disparities, we observe marked differences between MSD for men and women across the 29 phenotypes. Using recall MSD as a proxy for model bias we compare models by ranking phenotypes by MSD magnitude per gender  (Fig. \ref{fig:mse_recall}). The recall MSD can be interpreted per phenotype, such as in Crohn's, where a small male-biased MSD contradicts our TTD finding that women present symptoms earlier. In Crohn's, on average women present more symptoms than men (23.86 vs 22.74 for women/men). This might indicate that while women present the right conditions earlier to help detect a phenotype, they also present many other conditions that confuse the classifier, thus rendering a male-biased recall. 
Across all phenotypes, recall MSD was slightly skewed towards men (17 phenotypes), and those with less data (smaller circles) showed higher MSD magnitude (far left or far right) (Fig. \ref{fig:mse_recall}). 

\vspace{-4ex}
\section{Conclusion and Future Work}
\label{sec:concl}
\vspace{-5pt}
We propose simple metrics and experiments to explore TTD to investigate gender disparities. We find that across 29 phenotypes, women were far more likely to experience a longer TTD than men. Despite this fact, when training gender-agnostic classifiers to recognize disease, we find that a majority perform better for men than women. Furthermore, this persists across most time windows prior to diagnosis. 

Our analysis is limited to a single clinical site. We note however that we operate in the OHDSI framework, and the methodology is extensible to other OHDSI sites and validated phenotype definitions.

\acks{This work was supported by NLM awards R01 LM013043 and T15 LM007079.}

\bibliography{jmlr-sample}

\begin{thebibliography}{12}
\providecommand{\natexlab}[1]{#1}
\providecommand{\url}[1]{\texttt{#1}}
\expandafter\ifx\csname urlstyle\endcsname\relax
  \providecommand{\doi}[1]{doi: #1}\else
  \providecommand{\doi}{doi: \begingroup \urlstyle{rm}\Url}\fi

\bibitem[Chen et~al.(2018)Chen, Johansson, and Sontag]{chen2018classifier}
Irene Chen, Fredrik~D Johansson, and David Sontag.
\newblock Why is my classifier discriminatory?
\newblock In \emph{Advances in Neural Information Processing Systems}, pages
  3539--3550, 2018.

\bibitem[Chen et~al.(2020)Chen, Pierson, Rose, Joshi, Ferryman, and
  Ghassemi]{20chen}
Irene Chen, Emma Pierson, Sherri Rose, Shalmali Joshi, Kadija Ferryman, and
  Marzyeh Ghassemi.
\newblock Ethical machine learning in health care.
\newblock \emph{Annual Review of Biomedical Data Science}, 2020.

\bibitem[Greenwood et~al.(2018)Greenwood, Carnahan, and
  Huang]{greenwood2018patient}
Brad~N Greenwood, Seth Carnahan, and Laura Huang.
\newblock Patient--physician gender concordance and increased mortality among
  female heart attack patients.
\newblock \emph{Proceedings of the National Academy of Sciences}, 115\penalty0
  (34):\penalty0 8569--8574, 2018.

\bibitem[Launay et~al.(2016)Launay, Cohen, Bossuyt, Buekens, Deeks, Dye,
  Feltbower, Ferrari, Kramer, Leeflang, et~al.]{launay2016reporting}
Elise Launay, J{\'e}r{\'e}mie~F Cohen, Patrick~M Bossuyt, Pierre Buekens,
  Jonathan Deeks, Timothy Dye, Richard Feltbower, Andrea Ferrari, Michael
  Kramer, Mariska Leeflang, et~al.
\newblock Reporting studies on time to diagnosis: proposal of a guideline by an
  international panel (rest).
\newblock \emph{BMC medicine}, 14\penalty0 (1):\penalty0 146, 2016.

\bibitem[Mehta et~al.(2016)Mehta, Beckie, DeVon, Grines, Krumholz, Johnson,
  Lindley, Vaccarino, Wang, Watson, et~al.]{mehta2016}
Laxmi~S Mehta, Theresa~M Beckie, Holli~A DeVon, Cindy~L Grines, Harlan~M
  Krumholz, Michelle~N Johnson, Kathryn~J Lindley, Viola Vaccarino, Tracy~Y
  Wang, Karol~E Watson, et~al.
\newblock Acute myocardial infarction in women: a scientific statement from the
  american heart association.
\newblock \emph{Circulation}, 133\penalty0 (9):\penalty0 916--947, 2016.

\bibitem[OHDSI()]{OHDSI}
OHDSI.
\newblock Observational health data sciences and informatics studies.
\newblock
  \url{https://github.com/ohdsi-studies/Covid19CharacterizationCharybdis/tree/master/inst/cohort}.

\bibitem[Regitz-Zagrosek(2012)]{regitz-zagrosek2012}
Vera Regitz-Zagrosek.
\newblock Sex and gender differences in health: Science \& society series on
  sex and science.
\newblock \emph{EMBO reports}, 13\penalty0 (7):\penalty0 596--603, 2012.

\bibitem[Regitz-Zagrosek and Seeland(2013)]{seeland2013}
Vera Regitz-Zagrosek and Ute Seeland.
\newblock Sex and gender differences in clinical medicine.
\newblock In \emph{Sex and Gender Differences in Pharmacology}, pages 3--22.
  Springer, 2013.

\bibitem[Shaw et~al.(2006)Shaw, Merz, Pepine, Reis, Bittner, Kelsey, Olson,
  Johnson, Mankad, Sharaf, et~al.]{shaw2006}
Leslee Shaw, Noel Merz, Carl Pepine, Steven Reis, Vera Bittner, Sheryl Kelsey,
  Marian Olson, Delia Johnson, Sunil Mankad, Barry Sharaf, et~al.
\newblock Insights from the {NHLBI}-sponsored women’s ischemia syndrome
  evaluation {(WISE)} study: Part {I}: gender differences in traditional and
  novel risk factors, symptom evaluation, and gender-optimized diagnostic
  strategies.
\newblock \emph{Journal of the American College of Cardiology}, 47\penalty0 (3
  Supplement):\penalty0 S4--S20, 2006.

\bibitem[Westergaard et~al.(2019)Westergaard, Moseley, S{\o}rup, Baldi, and
  Brunak]{westergaard2019}
David Westergaard, Pope Moseley, Freja Karuna~Hemmingsen S{\o}rup, Pierre
  Baldi, and S{\o}ren Brunak.
\newblock Population-wide analysis of differences in disease progression
  patterns in men and women.
\newblock \emph{Nature communications}, 10\penalty0 (1):\penalty0 1--14, 2019.

\bibitem[{World Health Organization}(2002)]{WHO2002}
{World Health Organization}.
\newblock {WHO} gender policy: Integrating gender perspectives in the work of
  {WHO}.
\newblock \emph{Geneva: World Health Organization}, page~p6, 2002.

\bibitem[Zhang et~al.(2020)Zhang, Lu, Abdalla, McDermott, and
  Ghassemi]{zhang2020hurtful}
Haoran Zhang, Amy~X Lu, Mohamed Abdalla, Matthew McDermott, and Marzyeh
  Ghassemi.
\newblock Hurtful words: quantifying biases in clinical contextual word
  embeddings.
\newblock In \emph{Proceedings of the ACM Conference on Health, Inference, and
  Learning}, pages 110--120, 2020.

\end{thebibliography}
\newpage
\appendix

\section{Disease Classification}\label{app:classifier}

Given the phenotype definition, a corresponding cohort is determined, i.e., a set of patients that obtain a diagnosis for the disease of interest along with an time of official diagnosis. The cohort is split into an 80\% train and 20\% test sets, with the split maintaining the gender proportions prior to the split. Negative examples for the classifier are drawn by finding patients from other phenotypes that were within a 5 year age with the same gender (Table~\ref{tab:ap1}).

For each phenotype, we train a binary classifier using L2-penalized logistic regression. Features consist of all extracted coded conditions up to the diagnosis time, represented as one-hot encodings.

Fig.~\ref{fig:crohn_performance} shows the performance for Crohn's disease in particular over the different test sets. 



\begin{table}[hb]
\resizebox{\columnwidth}{!}{%
\begin{tabular}{l|cc}
\toprule
Phenotype & Train  &    Test    \\
\midrule 
Angioedema & 458 & 114 \\ 
Gastrointestinal bleeding & 4280 & 1070 \\ 
Gout & 2588 & 648 \\ 
Heart failure & 6388 & 1596 \\ 
Hepatic failure & 1444 & 362 \\ 
Prevalent Type 2 Diabetes & 28138 & 7036 \\ 
Prevalent hypertension & 61258 & 15314 \\ 
Prevalent chronic kidney disease & 8436 & 2110 \\ 
Prevalent ESRD & 2170 & 542 \\ 
Prevalent heart disease & 52938 & 13234 \\ 
Prevalent malignant neoplasm*  & 24638 & 6160 \\ 
Prevalent HIV infection & 1502 & 376 \\ 
Prevalent Dementia & 6956 & 1740 \\ 
Prevalent COPD without asthma & 14130 & 3532 \\ 
Prevalent Asthma without COPD & 22336 & 5584 \\ 
Acute myocardial infarction events & 3856 & 964 \\ 
Suicide and suicidal ideation & 4052 & 1014 \\ 
Psoriasis & 2364 & 592 \\ 
Psoriatic arthritis & 340 & 86 \\ 
Multiple sclerosis & 408 & 102 \\ 
Systemic lupus erythematosus & 432 & 108 \\ 
Sjogrens syndrome & 206 & 52 \\ 
Hashimoto thyroiditis & 418 & 104 \\ 
Myasthenia gravis & 246 & 62 \\ 
Celiac disease & 762 & 190 \\ 
Scleroderma & 214 & 54 \\ 
Sarcoidosis & 8352 & 2088 \\ 
Ulcerative colitis & 1128 & 284 \\ 
Crohn's disease & 5074 & 1270 \\ 
\bottomrule
\end{tabular}}
\caption{Number of patients in the training and test sets for the 29 phenotypes. *Excludes non-melanoma skin cancers.}
\label{tab:ap1}
\end{table}

\begin{figure}[thb]
\floatconts
    {fig:crohn_performance}
    {\caption{Performance of the Crohn's diagnosis classifier when tested with varying levels of right censoring.}}
    {\includegraphics[width=\columnwidth]{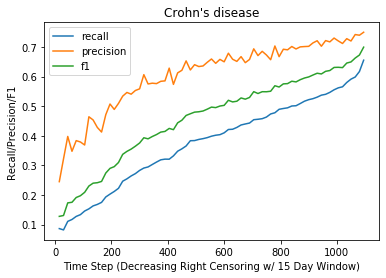}}
\end{figure}
\vfill\null
\section{Fairness Gap Metrics and Results}\label{apd:fairness}
The mean squared discrimination (MSD) can be expressed mathematically as $\text{MSD}(M_d) = \text{sign}(\frac{1}{b} \sum_{i=1}^b g_i) * (\frac{1}{b} \sum_{i=1}^b g_i^2)$, where $g_i$ is the fairness gap (Table \ref{tab:fairness_gaps}) at window $i$. The model's MSD is proportional to the mean squared error of recall, precision, or specificity gap, to penalize biases towards any gender and to heavily penalize models with larger biases; it is also sign dependent on the gap mean's sign to signify which gender the model exhibits better performance towards. The MSD is constrained between [-1, 1]; since fairness gaps are between [0, 1], the MSE will be between [0, 1], and the gender sign flips the value from -1 to 1. For gender, positive MSD indicates model bias favoring men, while negative MSD favors women. 

\begin{figure}[htb]
\floatconts
    {fig:crohn_recall_gap}
    {\caption{Gender-specific recall gaps at varying levels of right censoring for Crohn's Disease.}}
    {\includegraphics[width=\columnwidth]{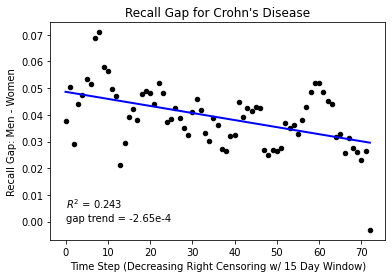}}
    \label{fig:crohn_recall_gap}
\end{figure}

Table~\ref{tab:fairness_gaps} shows the definition for the different gap metrics we use. 
Figure~\ref{fig:crohn_recall_gap} shows for Chron's disease in particular, a scatter plot of the recall gaps at varying levels of right censoring (men $-$ women). 

\begin{figure*}
\centering
\includegraphics[width=0.7\textwidth]{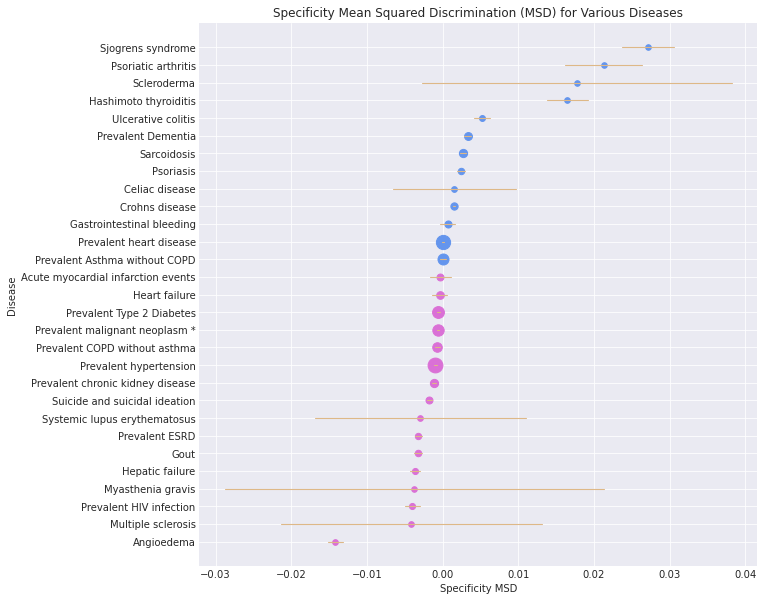}
\caption{Specificity MSD across phenotypes. Blue circles indicate better recall for men; pink, for women. Circle size is proportional to the size of the dataset (Table \ref{tab:ap1}). Error bars correspond to 95\% confidence intervals.  {\tiny *Excludes non-melanoma skin cancer  }}\label{fig:mse_specificity}
\end{figure*}

\begin{figure*}
\centering
\includegraphics[width=0.7\textwidth]{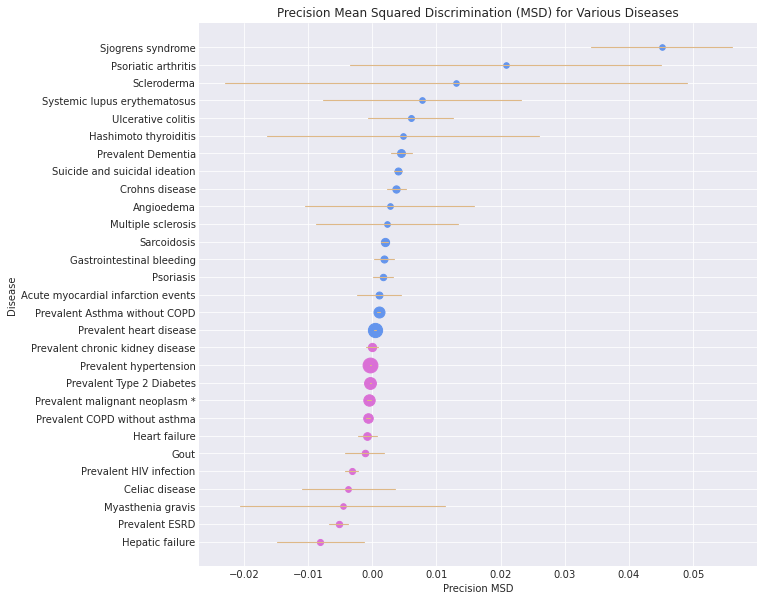}
\caption{Precision MSD across phenotypes. Blue circles indicate better recall for men; pink, for women. Circle size is proportional to the size of the dataset (Table \ref{tab:ap1}). Error bars correspond to 95\% confidence intervals.  {\tiny *Excludes non-melanoma skin cancer  }}\label{fig:mse_precision}
\end{figure*}

In addition to the mean recall bias across the 29 phenotypes, Figures~\ref{fig:mse_precision} and \ref{fig:mse_specificity} show the mean specificity bias and the mean precision gap.

\begin{table}[]
\resizebox{\columnwidth}{!}{%
\begin{tabular}{ll}
\toprule
Gap             & Definition \\ 
\midrule
Recall Gap      & $\dfrac{\text{TP}_1}{\text{TP}_1 + \text{FN}_1} - \dfrac{\text{TP}_2}{\text{TP}_2 + \text{FN}_2}$  \\[10pt]
Specificity Gap & $\dfrac{\text{TN}_1}{\text{TN}_1 + \text{FP}_1} - \dfrac{\text{TN}_2}{\text{TN}_2 + \text{FP}_2}$ \\[10pt]
Precision Gap   & $\dfrac{\text{TP}_1}{\text{TP}_1 + \text{FP}_1} - \dfrac{\text{TP}_2}{\text{TP}_2 + \text{FP}_2}$  \\[10pt] 
\bottomrule
\end{tabular}}
\caption{Fairness gap metrics used in time-variant model fairness analysis.}
\label{tab:fairness_gaps} 
\end{table}

\end{document}